\documentclass[aps,preprint]{revtex4}%
\usepackage{amsfonts}
\usepackage{amsmath}
\usepackage{amssymb}
\usepackage{graphicx}%
\begin{document}
\preprint{ }
\title{Weiss oscillations in the magnetoconductivity of modulated graphene bilayer }
\author{M. Tahir$^{\ast}$}
\affiliation{Department of Physics, University of Sargodha, Sargodha 40100, Pakistan}
\author{K. Sabeeh}
\affiliation{Department of Physics, Quaid-i-Azam University, Islamabad 45320, Pakistan}

\begin{abstract}
We present a theoretical study of Weiss oscillations in magnetoconductivity of
bilayer graphene. Bilayer graphene in the presence of a perpendicular magnetic
field and a unidirectional weak electric modulation is considered.We determine
the $\sigma_{yy}$ component of the magnetoconductivity tensor for this system
which is shown to exhibit Weiss oscillations. We show that Weiss oscillations
in the magnetoconductivity of bilayer graphene are enhanced and more robust
with temperature compared to those in conventional two-dimensional electron
gas systems whereas they are less robust with temperature compared to
monolayer graphene. In addition, we also find phase differences of $\pi$ and
$2\pi$ in the magnetoconductivity oscillations compared to monolayer graphene
and conventional 2DEG system which arises due to the chiral nature of
quasiparticles in graphene.

\end{abstract}
\startpage{01}
\endpage{02}
\maketitle

\section{Introduction}

The successful preparation of monolayer graphene has allowed the possibility
of studying the properties of electrons in graphene \cite{1}. The nature of
quasiparticles called Dirac electrons in these two-dimensional systems is very
different from those of the conventional two-dimensional electron gas (2DEG)
realized in semiconductor heterostructures. Graphene has a honeycomb lattice
of carbon atoms. The quasiparticles in graphene have a band structure in which
electron and hole bands touch at two points in the Brillouin zone. At these
Dirac points the quasiparticles obey the massless Dirac equation. In other
words, they behave as massless Dirac particles leading to a linear dispersion
relation $\epsilon_{k}=vk$ ( with the characteristic velocity $v\simeq
10^{6}m/s)$. This difference in the nature of the quasiparticles in graphene
from conventional 2DEG has given rise to a host of new and unusual phenomena
such as anamolous quantum Hall effects and a $\pi$ Berry phase\cite{1}%
\cite{2}. Besides the fundamental interest in understanding the electronic
properties of graphene there is also serious suggestions that it can serve as
the building block for nanoelectronic devices \cite{3}.

In addition to the graphene monolayer, there has been a lot of interest in
investigating the properties of bilayer graphene. The quasiparticles in
bilayer graphene exhibit a parabolic dispersion relation which implies that
they are massive particles. These quasiparticles are also chiral and are
described by spinor wavefunctions\cite{2,4,5,6,7}. Recent theoretical work on
graphene multilayers has also shown the existance of Dirac electrons with a
linear energy spectrum in monolayer graphene and a parabolic spectrum for
Dirac electrons in bilayer\cite{4}. Bilayer graphene consists of two
monolayers stacked as in natural graphite. This, Bernal stacking, yields a
unit cell of four atoms with the result that there are four electronic bands.
In $k$ space, the bilayer has a hexagonal Brillouin zone. Its physical
properties are mainly determined by the eigenvalues and eigenfunctions at two
inequivalent corners of the Brillouin zone, $K$ and $K^{\prime},$ where the
$\pi^{\ast}$ conduction and $\pi$ valence bands meet at the Fermi surface. Due
to the strong interlayer coupling both the conduction and valence bands are
split \ by an energy $\sim0.4eV$ near the $K$ and $K^{\prime}$ valleys. Only
two of these bands, upper valence and lower conduction band, are relevant at
low energy and they can be described by the effective Hamiltonian given
below\cite{2,5,6}

It was found years ago that if conventional 2DEG is subjected to artificially
created periodic potentials it leads to the appearence of Weiss oscillations
in the magnetoresistance. This type of electrical modulation of the 2D system
can be carried out by depositing an array of parallel metallic strips on the
surface or through two interfering laser beams \cite{8,9,10}. Weiss
oscillations were found to be the result of commensurability of the electron
cyclotron diameter at the Fermi energy and the period of the electric
modulation. These oscillations were found to be periodic in the inverse
magnetic field \cite{8,9,10}. Recently, an investigation of electric field
modulation effects on transport properties in monolayer graphene has led to
the prediction of enhanced Weiss oscillations in the
magnetoconductivity\cite{11}. In addition, the magnetoplasmons spectrum,
density of states, bandwidth and thermodynamics properties of monolayer
graphene in the presence of electrical modulation have been investigated so
far\cite{13}. In this work we are interested in studying the effects of
electric modulation on magnetoconductivity in bilayer graphene and we compare
the results obtained with those of monolayer graphene and the conventional 2DEG.

\section{Formulation}

We consider symmetric bilayer graphene within the single electron
approximation described by the following effective Hamiltonian ($\hbar=c=1$
here)\cite{2,5}%
\begin{equation}
H_{0}=-\frac{1}{2m}\left(
\begin{array}
[c]{c}%
0\text{ \ \ \ \ \ \ (}P_{x}-iP_{y}\text{)}^{2}\\
\text{(}P_{x}+iP_{y}\text{)}^{2}\text{ \ \ \ \ \ \ }0\text{ }%
\end{array}
\right)  ,\label{1}%
\end{equation}
where $\overrightarrow{p}=-i\overleftrightarrow{\nabla}-e\overleftrightarrow
{A}$, with the vector potential expressed in the Landau gauge as
$\overleftrightarrow{A}=(0,Bx,0)$ and the magnetic field is $B=\left(
0,0,B\check{z}\right)  ,$which is perpendicular to the bilayer graphene, $m$
is the effective mass of the electrons in bilayer: $m=0.043m_{e}$ with $m_{e}$
the usual electron mass. The energy eigenvalues and eigenfunction in the
presence of the magnetic field are%
\begin{equation}
\varepsilon(n)=\omega_{c}\sqrt{n(n-1)},\text{ \ \ }n\gtrsim2\label{2}%
\end{equation}
where $\omega_{c}=\frac{eB}{m}$ is the cyclotron frequency. For the low
magnetic fields considered in this work, the Hamiltonian of Eq.(1) and the
Landau level spectrum in Eq.(2) adequately captures the low energy electronic
properties in bilayer in the presence of a magnetic field\cite{5}. The
eigenfunction can be written as%
\begin{equation}
\Psi_{n,K_{y}}^{k}(r)=\frac{e^{iK_{y}}}{\sqrt{2L_{y}}}\left(
\begin{array}
[c]{c}%
\Phi_{n-2}\\
\Phi_{n}%
\end{array}
\right)  ,\label{3}%
\end{equation}
where $L_{y}$ is the $y-$dimension of the bilayer and the normalized harmonic
oscillator eigenfunction are%
\[
\Phi_{n}(x)=\frac{1}{\sqrt{2^{n}n\sqrt{\pi}l}}\exp^{[-\frac{1}{2}\left(
\frac{x-x_{0}}{l}\right)  ^{2}]}H_{n}(\frac{x+x_{0}}{l}),
\]
with center of the cyclotron orbit $x_{0}=l^{2}k_{y}.$ We now consider a weak
one-dimensional periodic electric modulation in the $x-$direction given by the
following Hamiltonian%
\begin{equation}
H^{\prime}=V_{0}\cos(Kx),\label{4}%
\end{equation}
where $K=2\pi/a$\ , $a$ is the period of modulation\ and $V_{0}$ is the
amplitude of modulation. We apply standard perturbation theory to determine
the first order correction to the unmodulated energy eigenvalues in the
presence of modulation with the result%
\begin{equation}
\varepsilon^{\prime}(n,x_{0})=V_{n}\cos(Kx_{0}),\label{5}%
\end{equation}
where%
\[
V_{n}(u)=\frac{V_{0}}{2}\exp(-u/2)(L_{n}(u)+L_{n-2}(u)),
\]
$u=K^{2}l^{2}/2,$and $L_{n}(u)$ are Laguerre polynomials.

From equations (2) and (5), the energy eigenvalues for the system in the
presence of modulation are%
\begin{equation}
\varepsilon(n,x_{0})=\omega_{c}\sqrt{n(n-1)}+V_{n}\cos(Kx_{0}). \label{6}%
\end{equation}
From equation (6) we observe that the formerly sharp Landau levels are now
broadened into minibands by the modulation potential. Furthermore, the Landau
bandwidth (\symbol{126}$\mid V_{n}\mid$) oscillate as a function of $n$, since
$L_{n}(u)$ is an oscillatory function of its index.

The bandwidth contains an average of Laguerre polynomials with indices $n$ and
$n-2$. To compare, in the electrically modulated monolayer graphene the
bandwidth depends on a linear combination of Laguerre polynomials with indices
$n$ and $n-1$ whereas for standard electrons in 2DEG there is only a single
term that contains Laguerre polynomial with index $n$. We expect that this
modulation induced change in the electronic density of states to influence the
magnetoconductivity of bilayer graphene and this is calculated in the
following section.

\section{Magnetoconductivity with Periodic Electric Modulation}

To determine the magnetoconductivity in the presence of weak electric
modulation we apply the Kubo formula in the linear response regime. In the
presence of the magnetic field, the main contribution to the Weiss
oscillations in magnetoconductivity arises from scattering induced migration
of the Larmor circle center. This is the diffusive conductivity and we shall
determine it following the approach in\cite{10,11,12}. In the case of
quasielastic scattering of the electrons, the diagonal component $\sigma_{yy}$
of the conductivity can be calculated by the following expression,%

\begin{equation}
\sigma_{yy}=\frac{\beta e^{2}}{L_{x}L_{y}}\underset{\zeta}{%
{\displaystyle\sum}
}f(\varepsilon_{\zeta})[1-f(\varepsilon_{\zeta})]\tau(\varepsilon_{\zeta
})(\upsilon_{y}^{\zeta})^{2} \label{7}%
\end{equation}
$L_{x}$, $L_{y}$, are the dimensions of the layer, $\beta=\frac{1}{k_{B}T}%
$\ is the inverse temperature with $k_{B}$ the Boltzmann constant,
$f(\varepsilon)$ is the Fermi Dirac distribution function and $\tau
(\varepsilon)$\ is the electron relaxation time and $\zeta$ denotes the
quantum numbers of the electron eigenstate.The diagonal component of the
conductivity $\sigma_{yy}$ is due to modulation induced broadening of Landau
bands and hence it carries the effects of modulation in which we are primarily
interested in this work. $\sigma_{xx}$ does not contribute as the component of
velocity in the $x$-direction is zero here. The collisional contribution due
to impurities is not taken into account in this work.

The summation in Eq.(7) over the quantum numbers $\zeta$ can be written as%
\begin{equation}
\underset{\zeta}{\frac{1}{A}%
{\displaystyle\sum}
}=\frac{L_{y}}{2\pi}%
{\displaystyle\int\limits_{0}^{\frac{L_{x}}{l^{2}}}}
dk_{y}\underset{n=0}{\overset{\infty}{%
{\displaystyle\sum}
}}=\frac{1}{2\pi l^{2}}\underset{n=0}{\overset{\infty}{%
{\displaystyle\sum}
}} \label{8}%
\end{equation}
where $A=L_{x}L_{y}$ is area of the system. The component of velocity required
in Eq.(7) can be calculated from the following expression%
\begin{equation}
\upsilon_{y}^{\zeta}=\frac{\partial}{\partial k_{y}}\varepsilon(n,x_{0}).
\label{9}%
\end{equation}
Substituting the expression for $\varepsilon(n,x_{0})$ obtained in Eq.(6) into
Eq.(9) yields
\begin{equation}
\upsilon_{y}^{\zeta}=\frac{2V_{n}(u)u}{K}\sin(Kx_{0}). \label{10}%
\end{equation}
With the results obtained in Eqs.(8), (9) and (10) we can express the
diffusive contribution to the conductivity given by Eq.(7) as%
\begin{equation}
\sigma_{yy}=A_{0}\phi\label{11}%
\end{equation}
where%
\begin{equation}
A_{0}=\frac{2}{\pi}V_{0}^{2}e^{2}\tau\beta\label{12}%
\end{equation}
and the dimensionless conductivity of bilayer graphene $\phi$ is given as
\begin{equation}
\phi=\frac{ue^{-u}}{4}%
{\displaystyle\sum_{n=0}^{\infty}}
\frac{g(\varepsilon(n))}{[g(\varepsilon(n))+1)]^{2}}[L_{n}(u)+L_{n-2}(u)]^{2}.
\label{13}%
\end{equation}
where $g(\varepsilon)=\exp[\beta(\varepsilon-\varepsilon_{F}]$ and
$\varepsilon_{F}$ is the Fermi energy.

\section{Asymptotic Expressions}

To get a better understanding of the results of the previous section we will
consider the asymptotic expression of conductivity where analytic results in
terms of elementary functions can be obtained\cite{11}. We shall compare the
asymptotic results for the dimensionless conductivity obtained in this section
with the results obtained for the electrically modulated conventional 2DEG
system. We shall also compare these results with recently obtained results for
the monolayer graphene that is subjected to only the electric modulation.

The asymptotic expression of dimensionless conductivity can be obtained by
using the following asymptotic expression for the Laguerre polynomials%
\begin{equation}
\exp^{-u/2}L_{n}(u)\rightarrow\frac{1}{\sqrt{\pi\sqrt{nu}}}\cos(2\sqrt
{nu}-\frac{\pi}{4}). \label{14}%
\end{equation}
Note that the asymptotic results are valid when many Landau Levels are filled.
We \ now take the continuum limit:%
\begin{equation}
n-->\frac{\varepsilon(n)}{\omega_{c}},\overset{\infty}{\underset{n=0}{%
{\displaystyle\sum}
}}-->%
{\displaystyle\int\limits_{0}^{\infty}}
\frac{d\varepsilon}{\omega_{c}} \label{15}%
\end{equation}
to express the dimensionless conductivity in Eq.(13) as the following integral%
\begin{equation}
\phi=\frac{1}{\pi}%
{\displaystyle\int\limits_{0}^{\infty}}
d\varepsilon\frac{g(\varepsilon)}{[g(\varepsilon)+1)]^{2}}\sqrt{\frac{u}{n}%
}\cos^{2}(\sqrt{u/n})\cos^{2}(2\sqrt{nu}-\frac{\pi}{4}) \label{16}%
\end{equation}
where $u=2\pi^{2}/b$ and the dimensionless magnetic field $b$ is introduced as
$b=\frac{B}{B^{\prime}}$ with $B^{\prime}=\frac{1}{ea^{2}}.$

Now assuming that the temperature is low such that $\beta^{-1}\ll
\varepsilon_{F}$ and replacing $\varepsilon=\varepsilon_{F}+s\beta^{-1}$, we
rewrite the above integral as%
\begin{equation}
\phi=\frac{\sqrt{2/\varepsilon_{F}b\omega_{c}}}{4\beta}\cos^{2}\left(
\frac{2\pi}{p}\right)
{\displaystyle\int\limits_{-\infty}^{\infty}}
\frac{4dse^{s}}{(e^{s}+1)^{2}}\cos^{2}\left(  \frac{2\pi p}{b}-\frac{\pi}%
{4}+\frac{4\pi}{p\omega_{c}}s\right)  \label{17}%
\end{equation}
where $p=k_{F}a=\sqrt{2\pi n_{e}}a$ is the dimensionless Fermi momentum of the
electron. To obtain an analytic solution we have also replaced $\varepsilon$
by $\varepsilon_{F}$ in the above integral except in the sine term in the integrand.

The above expression can be expressed as
\begin{equation}
\phi=\frac{\sqrt{2/\varepsilon_{F}b\omega_{c}}}{4\beta}\cos^{2}\left(
\frac{2\pi}{p}\right)
{\displaystyle\int\limits_{-\infty}^{\infty}}
\frac{ds}{\cosh^{2}(s/2)}\cos^{2}\left(  \frac{2\pi p}{b}-\frac{\pi}{4}%
+\frac{4\pi}{p\omega_{c}}s\right)  . \label{18}%
\end{equation}
The above integration can be performed by using the following identity
\begin{equation}%
{\displaystyle\int\limits_{0}^{\infty}}
dx\frac{\cos ax}{\cosh^{2}\beta x}=\frac{a\pi}{2\beta^{2}\sinh(a\pi/2\beta)}
\label{19}%
\end{equation}
with the result%
\begin{equation}
\phi=\frac{T}{4\pi^{2}T_{B}}\cos^{2}\left(  \frac{2\pi}{p}\right)  \left[
1-A\left(  \frac{T}{T_{B}}\right)  +2A\left(  \frac{T}{T_{B}}\right)  \cos
^{2}\left[  2\pi\left(  \frac{p}{b}-\frac{1}{8}\right)  \right]  \right]
\label{20}%
\end{equation}
where $T_{B}$\ is the characteristic damping temperature of Weiss oscillations
in bilayer graphene: $k_{B}T_{B}=\frac{bp}{4\pi^{2}ma^{2}},$ $\frac{T}{T_{B}%
}=\frac{4\pi^{2}ma^{2}}{bp}$ and $A(x)=\frac{x}{\sinh(x)}-^{(x-->\infty
)}->=2xe^{-x}.$

\section{Comparison with Electrically modulated monolayer graphene}

We will now compare the results obtained in this work with results obtained in
\cite{11} for the case of electrically modulated monolayer graphene system. We
will first compare the energy spectrum in the two cases. The difference in the
energy spectrum due to modulation effects was obtained in Eq.(6). If we
compare this result with the corresponding expression for the electrically
modulated monolayer graphene case, we find the following differences: Firstly,
in the monolayer we have an average of two successive Laguerre polynomials
with indices $n$ and $n-1$ whereas here we also have the average of two
Laguerre polynomials but not successive ones but rather with indices $n$ and
$n-2$. Secondly, in the monolayer the energy eigenvalues are multiplied by the
square root of the Landau band index $\sqrt{n}$ whereas in the bilayer we have
$\sqrt{n(n-1)}$ factor. Thirdly, the cyclotron frequency in the two systems is
different since the quasiparticles in monolayer are massless Dirac particles
whereas they have a finite mass in the bilayer. These differences cause the
velocity expression for the electrons given by Eq.(10) to be different in the
two systems.

We now compare the expressions for dimensionless conductivity $\phi$ given by
Eq. (20) with the electrically modulated case (Eq.(22) in \cite{11}). The
argument of the cosine terms in the expression for bilayer are 2$\pi/p$
whereas in monolayer it is $\pi/p$ which results in the phase difference of
$\pi$ in the the dimensionless conductivity in the two systems. This we expect
as the quasiparticles in graphene (both monolayer and bilayer) are chiral and
acquire a Berry's phase in the presence of a magnetic field\cite{1}. The
Berry's phase acquired by Dirac electrons in monolayer graphene is $\pi$
whereas it is $2\pi$ for particles in bilayer graphene\cite{2,5}. Therefore we
observe a difference in phase of $\pi$ in the magnetoconductivity oscillations
in the two systems. The dimensionless magnetoconductivity for both
electrically modulated mono- and bi-layer graphene as a function of inverse
magnetic field is shown in Fig.(1)at temperature $T=6K$ , electron density
$n_{e}=2.3\times10^{11}cm^{-2}$ and period of modulation $a=350nm$.We also
observe that in the region of high magnetic field SdH oscillations are
superimposed on the Weiss oscillations. The oscillations are periodic in $1/B$
and the period depends on electron density as $\sqrt{n_{e}}.$

\section{Comparison with standard electron system in 2DEG}

We start by comparing the energy spectrum and velocity expression obtained in
Eq.(6) and Eq.(10) with similar expressions for the conventional 2DEG where
the the quasiparticles are standard electrons \cite{9}. For the energy
spectrum, we find that the Landau level spectrum is significantly different
from that of standard electrons in conventional 2DEG. The first term
$\omega_{c}\sqrt{n(n-1)}$ in Eq.(6) has to be compared with $\omega
_{c}(n+1/2)$ with $\omega_{c}=eB/m_{e}$ for standard electrons. Not only the
dependence on the Landau level index $n$ is different in the two systems but
the cyclotron frequency is also not the same due to the difference in mass of
the quasiparticles. The modulation effects are carried by the second term
where the essential difference is in the structure of the function
$V_{n}(u)=\frac{V_{0}}{2}\exp(-u/2)(L_{n}(u)+L_{n-2}(u)).$ We find that there
is a basic difference: In bilayer we have a sum of two Laguerre polynomials
with indices $n$ and $n-2$ whereas only a single Laguerre polynomial appears
in the corresponding term for standard electrons in 2DEG. This difference in
the $V_{n}(u)$ function causes the velocity expression for the electrons in
bilayer given by Eq.(10) to be different from that of the standard electrons.
To highlight the difference in the dimensionless conductivity in the two
systems, we compare the asymptotic expression in bilayer Eq.(20) with the
corresponding expression for 2DEG (Eq. (25) in \cite{11}). We find that
dimensionless conductivity in bilayer has an additional prefactor $\cos
^{2}\left(  \frac{2\pi}{p}\right)  $ which is not present in the corresponding
expression for 2DEG. In addition, conductivity in bilayer contains the
characteristic damping temperature $T_{B}$ which is higher than the
corresponding damping temperature in 2DEG $T_{p}$ due to the smaller effective
mass of the quasiparticles in bilayer. This results in the magnetoconductivity
oscillations to be more robust with temperature than in 2DEG. To see the
effects of this difference on the magnetoconductivity we present the
dimensionless magnetoconductivity for both electrically modulated bilayer
graphene and the electrically modulated standard 2DEG in Fig.(2),as a function
of inverse magnetic field at temperature $T=6K$ , electron density
$n_{e}=2.3\times10^{11}cm^{-2}$ and period of modulation $a=350nm$. We find
that the there is a difference in phase of $2\pi$ between the oscillations in
magnetoconductivity in the two systems since the quasiparticles in bilayer
graphene are chiral. A Berry's phase of $2\pi$ is acquired by the
quasiparticles in bilayer relative to the standard electrons resulting in the
appearence of $2\pi$ phase difference in the magnetoconductivity oscillations.
We also find a peak in magnetoconductivity in 2DEG that is absent in bilayer
which is due to the absence of contribution from the $n=0$ and $n=1$ Landau
levels as they lie at zero energy.

We also find that the magnetoconductivity oscillations in bilayer graphene are
less damped by temperature and are more prounced as compared to those in
conventional 2DEG system whereas they are less pronounced and are more damped
with temperature compared to those in monolayer graphene. This can be seen in
Fig.(3) where dimensionless conductivity as a function of inverse magnetic
field is presented for the three systems. The parameters used are: $T=6K$ ,
electron density $n_{e}=2.3\times10^{11}cm^{-2}$ and period of modulation
$a=350nm$. This can be understood by considering the temperature scale for
damping of Weiss oscillations in bilayer graphene obtained from Eq.(20) which
is characterized by $T_{B}$ given as $k_{B}T_{B}=\frac{bp}{4\pi^{2}ma^{2}}$
whereas the characteristic tempererature for 2DEG is given in \cite{10,11} as
$k_{B}T_{p}=\frac{bp}{4\pi^{2}m_{e}a^{2}}.$ Comparing $T_{B}$ and $T_{p}$ the
essential difference is the difference in the effective masses of the
quasiparticles in the two systems. Since the quasiparticles in bilayer have a
smaller effective mass $m=0.043m_{e},$ the characteristic damping temperature
$T_{B}$ is higher in bilayer than in conventional 2DEG characterized by
$T_{p}$. Hence Weiss oscillations in magnetoconductivity in bilayer graphene
are less damped with temperature compared to 2DEG system.

\section{Conclusions}

We have investigated the diffusive magnetoconductivity component $\sigma_{yy}$
in bilayer graphene in the presence of a perpendicular magnetic field and a
one-dimensional weak electric modulation. In this work, we focus on the Weiss
oscillations in magnetoconductivity. We have compared the results obtained
with those of electrically modulated monolayer graphene as well as
electrically modulated conventional 2DEG system. We find phase differences of
$\pi$ and $2\pi$ in the magnetoconductivity oscillations compared to monolayer
graphene and conventional 2DEG system which arises due to the chiral nature of
quasiparticles in graphene.We also find that the oscillations due to
modulation in the magnetoconductivity are enhanced and less damped with
temperature compared to conventional 2DEG system whereas they are less robust
with temperature compared to monolayer graphene.

\section{Acknowledgements}

The authors gratefully acknowledge helpful discussions with F. M. Peeters
during the preparation of this manuscript. One of us (K.S.) would also like to
acknowledge the support of the Pakistan Science Foundation (PSF) through
project No. C-QU/Phys (129). M. T. would like to acknowledge the support of
the Pakistan Higher Education Commission (HEC).

$\ast$Present address: The Blackett Laboratory, Imperial College London, SW7
2AZ London, United Kingdom.

\end{document}